# Humanistic Approach in Mobile Adhoc Network: HAMANET

[1]Md. Amir Khusru Akhtar and [2]G. Sahoo

[1]Department of Computer Science and Engineering,
Cambridge Institute of Technology, Ranchi, Jharkhand, India

[2]Department of Information Technology,
Birla Institute of Technology, Mesra, Ranchi, Jharkhand, India

`[1]akru2008@gmail.com`, `[2]gsahoo@bitmesra.ac.in`

**Abstract.** Human society is a complex and most organized networks, in which many communities have different cultural livelihood. The creation/formation of one or more communities within a society and the way of associations can be mapped to MANET. By involving human characteristics and behavior, surely it would pave a new way, for further development. In this paper we have presented a new approach called "HAMANET" which is not only robust and secure but it certainly meets the challenges of MANET (such as name resolution, address allocation and authentication). Our object oriented design defines a service in terms of Arts, Culture, and Machine. The 'Art' is the smallest unit of work (defined as an interface), the 'Culture' is the integration/assembling of one or more Arts (defined as a class) and finally the 'Machine' which is an instance of a Culture that defines a service. The grouping of the communicable Machines of the same Culture forms a 'Community'. We have used the term 'Society' for MANET consisting of one or more communities and modeled using humanistic approach. We have compared our design with GloMoSim and proposed the implementation of file transfer service using the said approach. Our approach gives better results in terms of implementation of the basic services, security, reliability, throughput, extensibility, scalability etc.

**Abbreviations**

Community Table (CT), Society Table (ST), Machine Culture (MC), Machine Identification (MID), Community Identification (CID), Service Initiator (SI), Machine Culture Start (MCSTART), Machine Culture Join (MCJOIN)

# 1 Introduction

As one can access the MANETs essence in the society, its immense use, the prolonged scatterration and the future requirement the foremost and futuristic thoughts can be further redefined and exaggerated from the needs of the acclimating society. The work which has been done so far has been very curative but it is high time to amend the ways for an energetic means to create a new platform for more research. The bondages of this study should finally pave a new way, for further development and create a new awareness amongst the researchers. By the involvement of the existing legacy and incorporating the necessary demand new eras will surely arrive.

To understand the approach first of all we would like to discuss creation/formation of human communities. A human community is formed to meet some objective(s). We have one or more communities in a society and a person is attached with one or more communities to fulfill its needs. From these discussions we would like to map that a MANET either has single service to run or more than one service. In this model Community term represents a running service involving a group of nodes in MANET. A MANET can run one or more services, and different communities are created to provide more than one services such as for file service, name service we have file sharing Community, name service Community respectively.

The "HAMANET Model" mapped the human attributes and behavior into a device[1] to evade the bondages of implementation and surely succeed as we live. Human network is the complex topology that still survives even in the case of known/unknown incidents. By defining the layout we certainly feel that this model assumes a node as a person that belongs to a society and further it is associated with one or more communities. That is why this anatomy is required.

This approach employs 'Art' the smallest unit of work (defined as an interface), the 'Culture' is the integration/assembling of one or more Arts (defined as a class) and finally the 'Machine' that is an instance of a Culture and define a service. A node must run a Machine to avail a service. The grouping of Machine forms a 'Community' having the same Culture, which define a running service in the network. Nodes can start a service and intimate this information by sending a MCSTART packet. Nodes that want to use that service send its consent using MCJOIN packet. The grouping of multiple different Community forms a Society.

The Arts can be implemented in terms of fine grain up to coarse grain. As we move towards fine grain to coarse grain certainly it gives simplicity in design but smaller customization. Fine grain Arts are extensible than coarse grain. The object oriented implementation surely gives a robust, secured, extensible and reliable platform to execute an application in the MANET that meets all the challenges that we have.

The rest of the paper is organized as follows. Section 2 presents the assumption and basic primitives. Section 3 briefs the related work. Section 4 presents the HAMANET model. Section 5 gives an analysis of the Model. Section 6 shows the implementation of FTP using HAMANET approach. Section 7 presents the conclusion and ongoing work.

---

1 It represents electronic equipment such as Computer, PDA, Mobile phones etc.

## 2 Assumptions and basic primitives

This section outlines the assumptions and basic primitives that we make regarding our humanistic approach.

### 2.1 Elementary terminologies

We have used following terms in designing our model

Definition 1. *Art (A): it denotes the smallest unit of work module needed to define a Culture and can be implemented in terms of fine grain up to coarse grain.*

Definition 2. *Culture (C): it denotes the integration/assembling of Arts that is used to define a service.*

Definition 3. *Machine (M): it denotes an instance of a Culture that defines a service and it must be a member of a Community.*

Definition 4. *Community (C): it denotes a group of communicable Machines having the same Culture that provide a service to the Society.*

Definition 5. *Society (S): it denotes the MANET consisting of one or more communities and modeled using humanistic approach.*

### 2.2 Human/device common basic needs

To design a MANET using a humanistic approach first we have analyzed the human and device basic needs to identify the selfish threats and attacks, the way of grouping to avail its services and the need of association. We have defined the common basic needs to understand the fact that establishes a mapping between human and device as given in Table 1.

**Table 1.** Basic Needs

| Human Needs | Device Needs | Art |
|---|---|---|
| Food | Energy | Energy saving Arts can be defined using the existing algorithms that save battery life and bandwidth. |
| Cloth | Privacy | Privacy Arts can be defined by involving existing rules that fulfill privacy needs. |
| House | Security | Secures Arts can be defined using existing security algorithms. |

### 2.3 Need of association

The association is an agreement between groups of individuals/nodes that form a community to accomplish a purpose. Peoples are living in a society in which they have different communities for different needs. We are defining association in Society/MANET because nodes have common objectives.

### 2.3.1 Communities

It means a particular 'class' of peoples having the same objective. In this work we are mapping this concept to define a community in which a set of nodes work together to run a service.

### 2.3.2 Society

A 'Society' refers to all communities of people having diverse goals. We have defined the 'Society' consisting of one or more Community for providing one or more services.

## 3 Related work

The work which has been done so far has been very curative but it is high time to amend the ways for an energetic means to create a new platform for more research. The bondages of this study should finally pave a new way, for further development and create a new awareness amongst the researchers. By the involvement of the existing legacy and incorporating the necessary demand new eras will surely arrive.

To our knowledge, there is no such design that we have proposed and no previously published work that meets all the challenges that we have in Mobile ad hoc networks. Lots of work has been proposed to handle modification of routing information's [3-7], to enforce cooperation [8] but they have serious limitations in terms of routing overhead and attacks. Work to be done in the area of Address allocation [17], name resolution [18, 19] shows that the implementation of these services with the current design of MANET creates complexity and heavy network overhead.

## 4 HAMANET Model

MANETs are certainly the next generation face that announces the game, winning which we obtain lave. The known/unknown obstacles resist the implementation, and by solving are circumventing to utilize the resources. The aggregated/generalized form of MANET can further specialize to esteem the demand that is why we are going to restructure the same and prevail. Human nature would really act into the device to entrust and make it robust and secure like the same.

### 4.1 Overview

Our object-oriented design defines a service in terms of Arts, Culture, and Machine. This approach employs 'Art' the smallest unit of work (defined as an interface), the 'Culture' is the integration/assembling of one or more Arts (defined as a class) and finally the Machine that is an instance of a Culture and define a service. A node must run a Machine to avail a service. The Machine defines a running service in a node, and the grouping of Machine having the same Culture forms a 'Community' defined in section 4.4.

We can provide the basic services in a MANET by instantiating a Machine after fulfilling the basic requirements for some services (for ex. Internet gateway point for name service). A SI node broadcast MCSTART packet in the Society/MANET to intimate that a service is being started. To join the service a node after receiving the MCSTART must send a MCJOIN packet to become a member of the Community or it can broadcast a MCJOIN packet to register itself into an existing Community. The lists of interested nodes that have replied by sending MCJOIN message are listed by the SI to form a Community.

In this model a community table is used in each Community that stores CID, MID and path. Packets are routed using the path provided in the community table. During transmission if a node doesn't find a route in the community table then it broadcasts a RREQ that is forwarded from node to node either node is a member or if in is not a member it handle it as friend RREQ. When the RREQ packet reaches to Community members, the members reply the source by sending the community table. This model defines levels of abstraction to handle malicious or selfish attacks as defined in section 4.5.

This will certainly give the platform upon which we can run and use any application in MANET. Fig. 1 shows the MANET that derives from a humanistic approach in which devices are defined using human behavior.

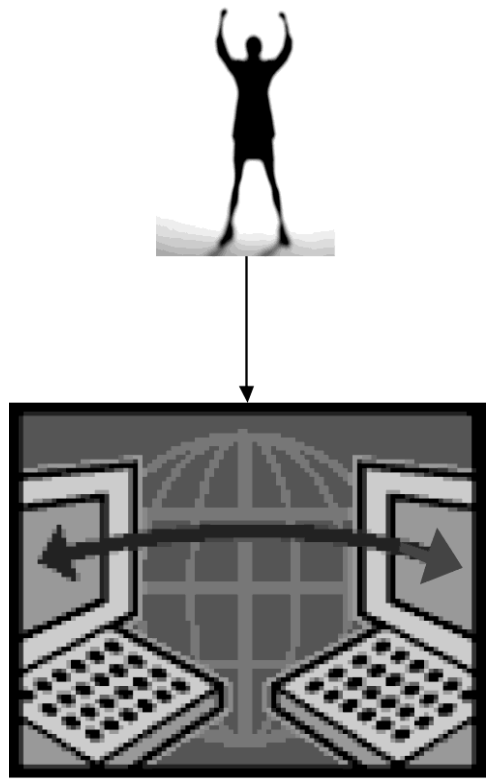

**Fig. 1.** Humanistic MANET

Humanistic approach defines the concept that in a society we have different community and each community is for different purposes/services. Similarly in a network we can create many communities with the same set of nodes to run many services. For example, a node transfers a file using Community 1, and logged on onto a remote node using Community 2.

### 4.2 HAMANET Arts/Cultures/Machines implementation Diagram

A Machine is an instance of a Culture and Culture implements Arts to define a service. Fig. 2 shows the implementation diagram in which we have defined Machines using Arts and Cultures.

where        Arts ranges from (1, 2, …, i)

                Cultures ranges from (1, 2, …, j)

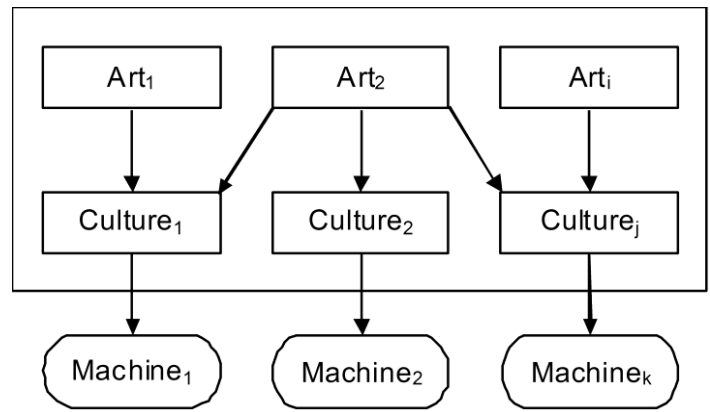

**Fig. 2.** Diagram showing interface (Arts), Class (Cultures), and object (Machines) in which a Culture is defined using Arts and finally instantiated to form a Machine

### 4.3 HAMANET Architecture

In this section we have defined the types of Society/MANET

#### 4.3.1 Single Community MANET

A 'single Community Society' is a MANET in which only one service is running such as file service, and it is created in the same way as defined in section 5.2. Fig. 3 shows a single Community MANET.

#### 4.3.2 Multiple Communities MANET

A 'multiple Communities Society' is defined as a MANET in which more than one services are running for example file service, name service etc. It is created in the same way as defined in section 5.2. Fig. 4 shows a multiple Community MANET.

In the figure we have used the notation M (MC, MID)
Where      M is the Machine
                MC is the Machine Culture
                MID is the Machine Identity within the Culture

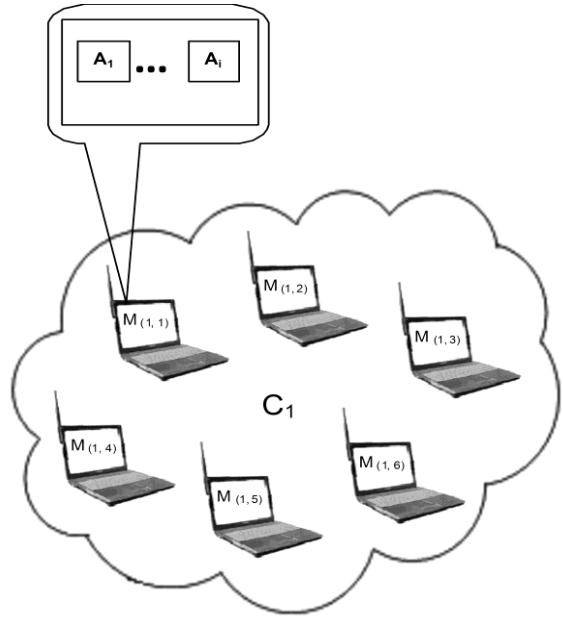

**Fig. 3.** Single Community MANET

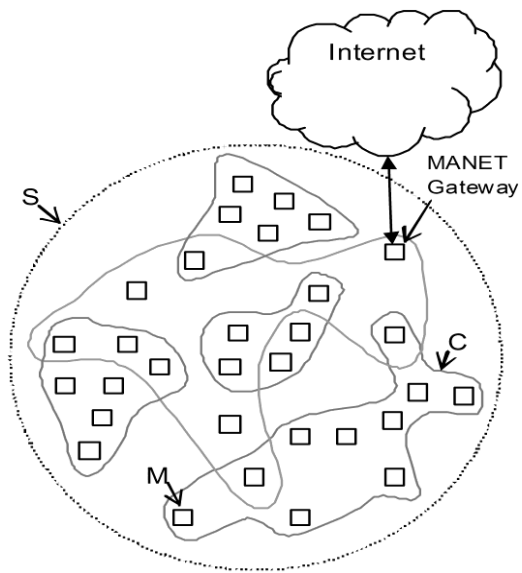

**Fig. 4.** A Society consisting of 5 communities

### 4.4 Description of the Components

The HAMANET components are as follows

#### 4.4.1 Arts Component

It defines the smallest work module(s) that is used to define a service and implemented in Culture. Art(s) can be defined in terms of fine grain up to coarse grain. As we move towards fine grain to coarse grain certainly it gives simplicity in design but little customization. The fine grain Art(s) can be defined at the function level (such as RREQ, RREP, RERR, HELLO etc.), similarly coarse grain Art(s) can be defined at the protocol level (such as Bellman-Ford, DSDV, FSR, OSPF, WRP, LAR, DSR, AODV) [9-16]. The Arts are defined in terms of interfaces that contain final variables and abstract methods defined in Table 2.

**Table 2.** Art Interface

| Interface Art1 |
| --- |
| final variables; |
| abstract methods (); |

#### 4.4.2 Culture Component

We have used Culture to assemble Arts and defined in terms of a class as given in Table 3. Cultures implements one or more Arts and instantiated to form a Machine.

**Table 3.** Culture Class

| Class Culture1 implements Art1, Art2, …, Arti |
| --- |
| Culture variables; |
| abstract methods (); // implementation of Arts interfaces |
| Culture methods (); |

#### 4.4.3 Machine Component and packets

A Machine is an instance of a Culture that defines a running service. To regularize its service a machine uses various functions that are defined in Culture methods (), they are as follows.

MCSTART: Packet broadcasted to the Society/MANET to intimate that a service has been started.

MCJOIN: Packet sends by interested Society/MANET member in response of MCSTART to join the Community.

### 4.4.4 Community Component

Community means the grouping of two or more Machines having the same Culture and communicable either directly or indirectly. Each Community maintains a community table listing all the members of the Community. The community table for node N1 is defined in Table 4.

**Table 4.** Community Table

| MID | CID | PATH |
|-----|-----|----------|
| N2  | C1  | N1-N2    |
| N3  | C1  | N1-N3    |
| N4  | C1  | N1-N2-N4 |

### 4.4.5 Society Component

All Communities finally constitute a Society that is defined for diverse goals. The Society Table is defined in Table 5.

**Table 5.** Society Table

| CID | MC |
|-----|--------------|
| C1  | File service |
| C2  | Name Service |

## 4.5 Levels of Abstraction

The communication between Machine object hides the internal details. The description of data at the object level is in a format independent of its Arts and Culture representation. Our work defines various levels of abstraction to minimize the complexity and protect the system from selfish and malicious threats as is given in Fig. 5.

Our object oriented design confirms that only defined operations can be executed with nodes because data's are wrapped by operations, which protects our network from malicious and selfish attacks.

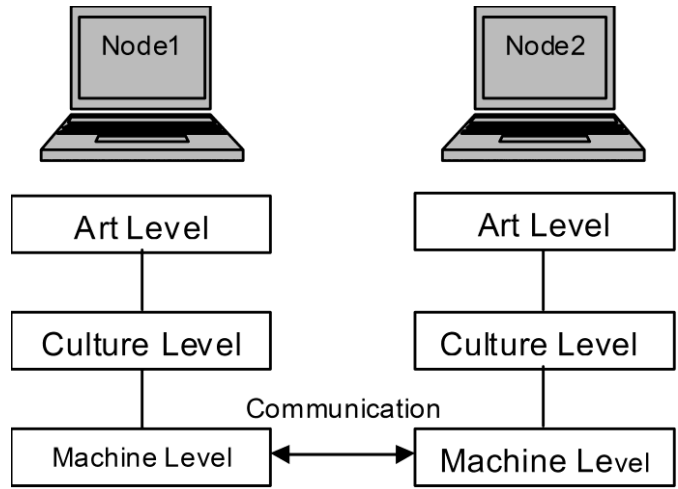

**Fig. 5.** Levels of Abstraction

## 5 Analysis of the Model

In this section we are describing the comparison and implementation of basic services.

### 5.1 Comparison between HAMANET Structure & GloMoSim Layers

The GloMoSim Layers [2] and the HAMANET Structure is given in Table 6 and Table 7 respectively. Through these tables we are comparing our proposed architecture with GloMoSim architecture.

**Table 6.** Layered Structure of GloMoSim

| Layer | Protocol |
|---|---|
| Physical (Radio Propagation) | Free space, Two-Ray |
| Data Link (MAC) | CSMA, MACA, TSMA, 802.11 |
| Network (Routing) | Bellman-Ford, FSR, OSPF, DSR, WRP, LAR, AODV |
| Transport | TCP, UDP |
| Application | Telnet, FTP |

Table 7.  HAMANET Structure

| Arts | | | | |
|---|---|---|---|---|
| Physical (Radio Propagation) | Data Link (MAC) | Network (Routing) | Transport | Application |
| Free space Two-Ray | CSMA MACA TSMA 802.11 | Bellman-Ford OSPF FSR DSR WRP LAR AODV | TCP UDP | Telnet FTP |

| Cultures | | |
|---|---|---|
| Culture 1 | Culture 2 | Culture 3 |
| Free space CSMA Bellman-Ford TCP FTP | Two-Ray 802.11 DSR TCP CBR | Two-Ray CSMA AODV TCP Telnet |

### 5.2 Implementing Basic Services

The Aleksi Penttinen [1] defined various challenges and services such as Address allocation, name resolution, authentication. Work to be done in the area of Address allocation, name resolution [17, 19] shows that the implementation of these services with the current design of MANET creates complexity and heavy network overhead. These services can be implemented using HAMANET without such problems.

To explain the implementation of these services first we relate our concept with creation of community in human. For example to create a community or club for some specific activity, two or more persons joined and make a community and then they add interested members and avail the service.

#### 5.2.1 Working Diagram

Our model is simpler than the existing MANET design in which a new service can be easily deployed. We simply create a list of Arts needed for implementing a service such as name service and assemble these Arts into the Culture and instantiate to form a Machine. The SI node broadcast MCSTART packet in the Society/MANET to intimate that a service is being started. To join the service a node after receiving the MCSTART must send a MCJOIN packet to become a member of the Community.

The lists of interested nodes that have replied by sending MCJOIN message are listed by the SI to form a Community. The working diagram for implementing basic services is given in Fig. 6. However, any of the nodes in the MANET will be connected to the Internet to provide the gateway that will be used to resolve the name.

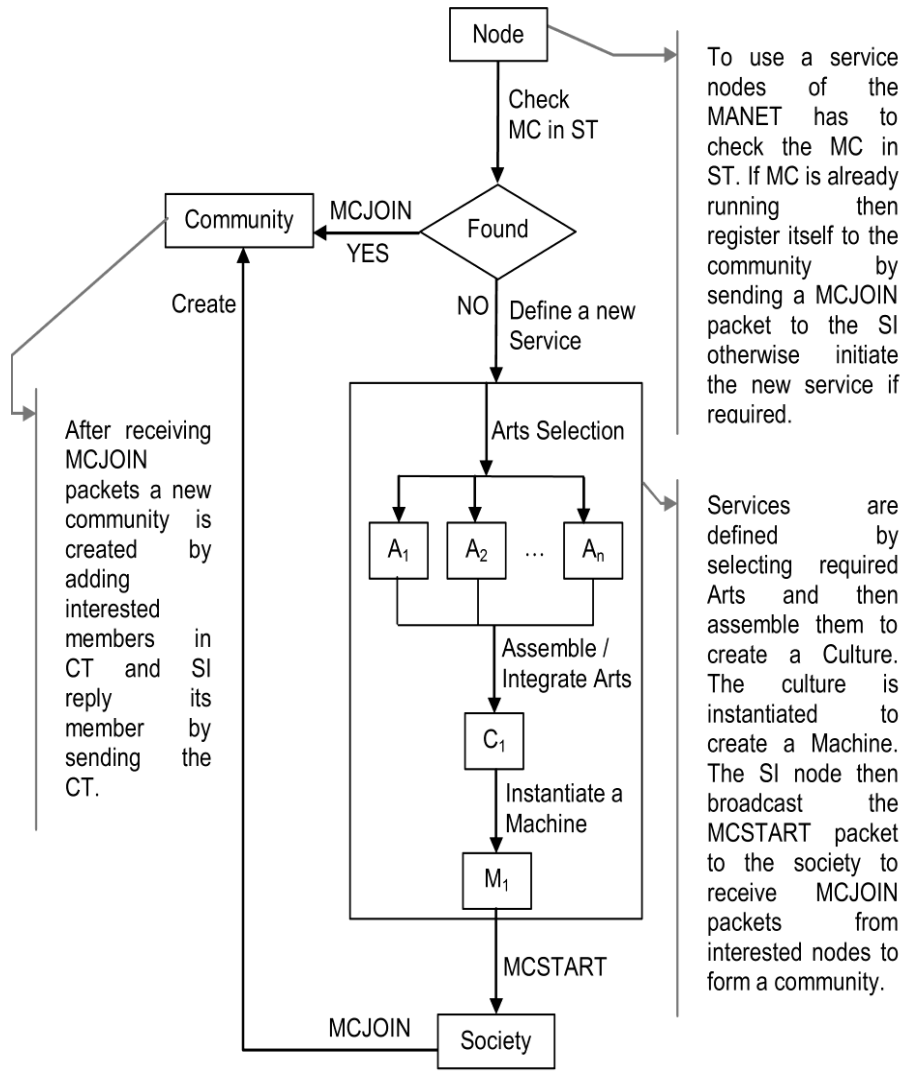

**Fig. 6.** Working model for initiating Basic services

### 5.2.2 Friend packets

If a node does not find a route in the community table it send friend packet to its reachable host to forward its packet to the Community member and the receiving host relay packets to the Community member.

### 5.2.3 Overhead

In our approach we were sending only one broadcast message at the time when a service is started to create a Community. After that we simply multicast messages within the Community to regularize and avail the services, this will degrade the message overhead in the MANET.

## 6 Case study: File transfer service

In this section we are proposing how File transfer service will be implemented using HAMANET. The protocol stack (Arts) needed to implement File transfer service is defined in Table 8.

**Table 8.** Protocol stack (Arts)

| Culture F |
|---|
| Free space |
| CSMA |
| DSDV |
| TCP |
| FTP |

The working model for implementation of file transfer service is similar to the implementation of basic services as defined in Fig. 6. This approach minimizes unnecessary broadcast due to regular update of its routing tables because it divides the MANET into more than one Community. In a network, participation varies from node to node, therefore a Community contains a list of interested members and thereby reduces the participation of idle nodes. This approach saves battery power and bandwidth as well as it is secured because it allows object communication.

## 7 Conclusion and ongoing work

In this paper we have presented a new approach called "HAMANET" that is robust and secure due to its object oriented design as well as because of humanistic approach it surely meet the challenges of the MANET. We have discussed the basic services and presented how these services can be implemented using the said approach. In the case study we have proposed File transfer service using our humanistic approach and it justifies, that it is better than the traditional implementation in terms of overhead and attacks.

This paper presents initial work on the humanistic MANET design. The long-term goal of this research project is to implement all the know-how in the MANET and to design a humanistic network simulator that fulfills the basics of the MANET.